\documentclass[twocolumn]{aastex631}
\usepackage{amsmath}

\graphicspath{{./}{pics/}}

\begin{document} 

\title{Sun-as-a-star Study of an X-class Solar Flare with Spectroscopic Observations of CHASE}

\author{Y. L. Ma}
\affiliation{School of Physics, Nanjing University, Nanjing 210023, People's Republic of China}
\author{Q. H. Lao}
\affiliation{School of Astronomy and Space Science, Nanjing University, Nanjing 210023, People's Republic of China; xincheng@nju.edu.cn}
\author{X. Cheng}
\affiliation{School of Astronomy and Space Science, Nanjing University, Nanjing 210023, People's Republic of China; xincheng@nju.edu.cn}
\affiliation{Key Laboratory of Modern Astronomy and Astrophysics (Nanjing University), Ministry of Education, Nanjing 210093, People's Republic of China}
\author{B. T. Wang}
\affiliation{School of Astronomy and Space Science, Nanjing University, Nanjing 210023, People's Republic of China; xincheng@nju.edu.cn}
\affiliation{Key Laboratory of Modern Astronomy and Astrophysics (Nanjing University), Ministry of Education, Nanjing 210093, People's Republic of China}
\author{Z. H. Zhao}
\affiliation{School of Astronomy and Space Science, Nanjing University, Nanjing 210023, People's Republic of China; xincheng@nju.edu.cn}
\affiliation{Key Laboratory of Modern Astronomy and Astrophysics (Nanjing University), Ministry of Education, Nanjing 210093, People's Republic of China}
\author{S. H. Rao}
\affiliation{School of Astronomy and Space Science, Nanjing University, Nanjing 210023, People's Republic of China; xincheng@nju.edu.cn}
\affiliation{Key Laboratory of Modern Astronomy and Astrophysics (Nanjing University), Ministry of Education, Nanjing 210093, People's Republic of China}
\author{C. Li}
\affiliation{School of Astronomy and Space Science, Nanjing University, Nanjing 210023, People's Republic of China; xincheng@nju.edu.cn}
\affiliation{Key Laboratory of Modern Astronomy and Astrophysics (Nanjing University), Ministry of Education, Nanjing 210093, People's Republic of China}
\affiliation{Institute of Science and Technology for Deep Space Exploration, Suzhou Campus, Nanjing University, Suzhou 215163, People's Republic of China}
\author{M. D. Ding}
\affiliation{School of Astronomy and Space Science, Nanjing University, Nanjing 210023, People's Republic of China; xincheng@nju.edu.cn}
\affiliation{Key Laboratory of Modern Astronomy and Astrophysics (Nanjing University), Ministry of Education, Nanjing 210093, People's Republic of China}

\begin{abstract}
Sun-as-a-star spectroscopic characteristics of solar flares can be used as a benchmark for the detection and analyses of stellar flares. Here, we study the Sun-as-a-star properties of an X1.0 solar flare using high-resolution spectroscopic data obtained by the Chinese $\mathrm{H} \alpha$ Solar Explorer (CHASE). A noise reduction algorithm based on discrete Fourier transformation is first employed to enhance the signal-to-noise ratio of the space-integral $\mathrm{H} \alpha$ spectrum with a focus on its typical characteristics. For the flare of interest, we find that the average $\mathrm{H} \alpha$ profile displays a strong emission at the line center and an obvious line broadening. It also presents a clear red asymmetry, corresponding to a redshift velocity of around $50 \ \mathrm{km \ s^{-1}}$ that slightly decreases with time, consistent with previous results. Furthermore, we study how the size of the space-integral region affects the characteristics of the flare Sun-as-a-star $\mathrm{H} \alpha$ profile. It is found that although the redshift velocity calculated from the $\mathrm{H} \alpha$ profile remains unchanged, the detectability of the characteristics weakens as the space-integral region becomes large. An upper limit for the size of the target region where the red asymmetry is detectable is estimated. It is also found that the intensity in $\mathrm{H} \alpha$ profiles, measured by the equivalent widths of the spectra, are significantly underestimated if the $\mathrm{H} \alpha$ spectra are further averaged in the time domain. 
\end{abstract}


\section{Introduction} \label{sec:intro}
Solar flares are one of the most powerful phenomena in the solar atmosphere, releasing a colossal amount of energy up to $\sim 10^{32} \ \mathrm{erg}$ \citep{RN52}. Researchers have made significant strides in interpreting how magnetic energy is converted into kinetic and thermal energy during a solar flare, which is related to magnetic reconnection. Observations at multi-passbands have played a crucial role in understanding the fundamental physical processes involved in solar flares, including the buildup, release, and transport of the energy \citep{RN53}. It should be noted that flares are frequently accompanied with coronal mass ejections (CMEs) \citep{RN54}, which can lead to severe space weather events such as geomagnetic storms, ionospheric storms, and variations in the density of Earth's upper atmosphere \citep{RN57}. 

Correspondingly, stellar flares have been studied and compared with solar flares in recent years. Various missions such as the Kepler Space Telescope \citep{Maehara_2012, Wu_2015, Davenport_2016, RN58} and the Transiting Exoplanet Survey Satellite (TESS) \citep{Doyle_2020} have observed tens of thousands of superflares that emit energy over $10^{33} \ \mathrm{erg}$ on stars. The relationship between chromospheric activity and stellar flares has been carefully studied with the aid of spectral data from the Chandra X-ray Observatory and the LAMOST telescope\citep{Lu_2019, Chen_2022, 2024ApJ...961..130Z}. It is believed that solar white-light flares and stellar superflares share similar energy release mechanisms, though the latter is characterized by a stronger magnetic field and a longer duration \citep{Namekata_2017}.
In addition, the evidence for chromospheric evaporation during stellar flares has been found \citep[e.g.,][]{Chen_2022}. Notably, both solar and stellar data align with a common scaling law, with the latter emitting hundreds of times more energy at the same peak temperature \citep{Aschwanden_2008}. Similar to solar flares, stellar flares can significantly impact the surrounding planets, including their chemical composition, climate, and dynamics \citep{RN56}. 
In light of these findings, the study of stellar flares has become a focal point of research, with significant implications for various fields, including solar and stellar physics, high-energy astrophysics, and planetary sciences.

However, little is known about the origin and evolution of stellar flares due to limited observations \citep{AlvaradoGomez2018SuppressionOC}. The lack of high spectral and temporal resolution data makes it difficult to study the physical processes during eruptive events, which are essential for understanding stellar atmospheric activities \citep{Spina_2020, wollmann2023observations, Pietrow2023}. Besides, it is also hard to distinguish among the major activities via observations without spatial resolution \citep{Vida_2019, Muheki_2020, Koller_2021, Maehara_2021}. Hence, theoretical analyses of stellar flare/CMEs and other related topics have been hindered by great uncertainties \citep{Lynch_2023}. 

To overcome the above difficulties, high-resolution solar data can be applied as a comparison to understand the physical mechanisms of flares on stars. To be specific, solar data can be spatially integrated to simulate stellar data and used to study the characteristics of various activities, and this way of analysis is named “Sun-as-a-star Analysis” \citep{livingston2007sun, RN59}. Sun-as-a-star extreme ultraviolet (EUV) observations have been used to study the correlations between flares and CMEs \citep{RN59, RN61, RN64} as well as to determine the line-of-sight (LOS) velocity of CMEs \citep{Brown_2016, RN60, Lu_2023}. In a study of 42 X-class solar flares, \citet{RN59} found that CMEs are often accompanied by coronal dimmings that usually appear at EUV passbands, which was later confirmed by \citet{RN61}. A study of the Doppler speeds of the hydrogen Lyman lines showed that upflows are associated with some kind of eruptions or coronal flows \citep{Brown_2016}. \citet{Lu_2023} then demonstrated the possibility of detecting stellar CMEs through blue asymmetries or blueshifts of spectral lines. In addition, by investigating the behavior of various photospheric and chromospheric spectral lines (including $\mathrm{H} \alpha$), \citet{Pietrow2023} concluded a relationship between the contrast profiles of flares over time and their locations. It is worth noting that $\mathrm{H} \alpha$ data, compared with EUV measurements, might be more important for the analyses of stellar activities since $\mathrm{H} \alpha$ radiation is much less influenced by interstellar absorption \citep{1994AJ107.2108R}. Blue- or red-asymmetry and line broadening in $\mathrm{H} \alpha$ line have been commonly identified in flares \citep{Namekata_2020, Koller_2021, Maehara_2021}. \citet{RN62} performed a Sun-as-a-star analysis on the solar $\mathrm{H} \alpha$ spectrum of an M-class flare utilizing data from the Solar Dynamics Doppler Imager (SDDI) and found signs of chromospheric condensation, namely red asymmetry and line broadening in the Sun-as-a star spectrum. Using a similar method, they then detected a potential eruptive filament from a stellar superflare by comparing the blueshifted $\mathrm{H} \alpha$ absorption component in the spectrum with solar flare profiles \citep{Namekata_2022}. In a study of 9 solar active events observed by SDDI, including several flares, \citet{RN63} concluded that filament eruptions show emission near the $\mathrm{H} \alpha$ line center with blue/red-shifted absorptions, and eruptions of off-limb prominences show blue or red-shifted emissions. 

The $\mathrm{H} \alpha$ Imaging Spectrograph (HIS) onboard the Chinese $\mathrm{H} \alpha$ Solar Explorer (CHASE) can acquire solar spectra near $\mathrm{H} \alpha \ (6559.7 - 6565.9 \ \mathrm{\text{\AA}})$ and $\mathrm{Fe \ I} \ (6567.8-6570.6 \ \mathrm{\text{\AA}})$ bands. Observational data from CHASE features a higher resolution than SDDI, specifically, 0.52" in pixel resolution and $0.024 \ \mathrm{\text{\AA}}$ in spectral sampling \citep{RN64}, enabling the recognition of more details of eruptive events. It therefore effectively facilitates the study of the evolution of integrated $\mathrm{H} \alpha$ profiles of solar activities and spectral comparisons between solar and stellar events. 

In this paper, we use the data from CHASE to perform a Sun-as-a-star analysis for an X-class flare to search for detailed spectroscopic characteristics and the correlations between those characteristics and the size of the integral area. Based on the Sun-as-a-star analyses method of \citet{RN63}, we introduce new calibration, noise reduction and fitting methods. Section \ref{sec:Method} gives a brief introduction to our target event and Sun-as-a-star integration methods. In Section \ref{sec:Result} we show our denoise algorithm and fitting methods as well as the corresponding results. Finally, in Section \ref{sec:Summary} we summarize our main findings, which are then followed by discussions. 

\section{Event and Methods} \label{sec:Method}
\subsection{Overview of the eruptive event} \label{subsec:Method 1}
\begin{figure*}[ht!]
{\centering\includegraphics[width=1.0\textwidth]{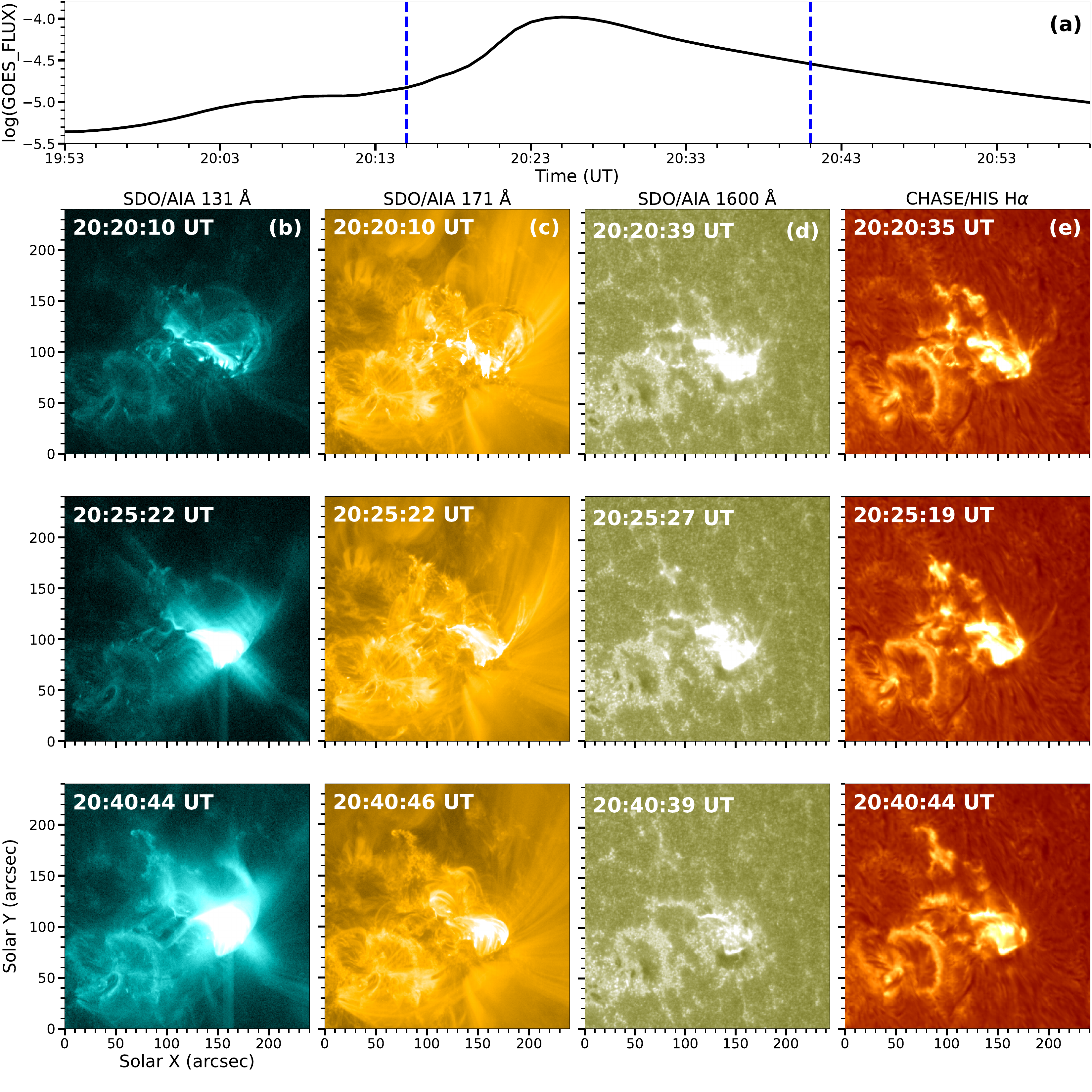}}
\caption{Overview of the target flare. 
(a) The GOES $1-8 \ \mathrm{\text{\AA}}$ SXR flux throughout the flare. CHASE observation time range is between the two vertical lines in blue. 
(b) - (e) SDO/AIA $131 \ \mathrm{\text{\AA}}$, $171 \ \mathrm{\text{\AA}}$, $1600 \ \mathrm{\text{\AA}}$ images and CHASE $\mathrm{H} \alpha$ line center images showing the temporal evolution of the $\mathrm{X1.0}$ flare.} 
\label{fig:1}
\end{figure*}

Our target event is an X1.0 flare that took place in the National Oceanic and Atmospheric Administration (NOAA) active region (AR) 13110 on October 2, 2022. Its GOES SXR flux reached its peak at $\text{20:25} \ \mathrm{UT}$ (Figure \ref{fig:1}a) and ended at $\text{20:34} \ \mathrm{UT}$. After the GOES peak time, a cusp-like structure was observed in the SDO/AIA images (Figure \ref{fig:1}b, $\text{20:40:44} \ \mathrm{UT}$, Figure \ref{fig:1}c, $\text{20:40:46} \ \mathrm{UT}$). 
This event is selected as the target event for two significant merits: (\romannumeral1) It is one of the most prominent events observed by CHASE, and (\romannumeral2) CHASE captured a part of the rising phase and especially the peak time of the flare (Figure \ref{fig:1}a). 

\subsection{CHASE/HIS Data Analysis} \label{subsec:Method 2}
Our study mainly uses the data provided by CHASE/HIS. We first explain the calibration process and the Sun-as-a-star method performed on the data. 

For raw data taken by HIS, we perform the dark-field correction, slit-image-curvature correction, flat-field correction, wavelength calibration, and coordinate transformation on it, thus obtaining Level 1 science data that can be used for further research \citep{RN67}. The CHASE/HIS Level 1 science data are available from the Solar Science Data Center of Nanjing University (SSDC) \footnote{https://ssdc.nju.edu.cn}.

Once acquire Level 1 $\mathrm{H} \alpha$ data, we implement a Sun-as-a-star analysis on them. 
It has been made clear in previous study \citep{RN63} that, even though Sun-as-a-star analysis stands for the spatial integration of the $\mathrm{H} \alpha$ spectra over the whole solar disk, the contribution of the solar flare region to the full disk radiation is relatively small compared to the entire non-flare region. To enhance the signals, we should only spatially integrate a relatively small region, namely a target region (TR), which contains the solar flare. 
The integration result is then averaged to one pixel and a contrast profile is calculated to simulate stellar observations. 
The details of our method are listed below. 

\begin{figure*}[ht!]
{\centering\includegraphics[width=1.0\textwidth]{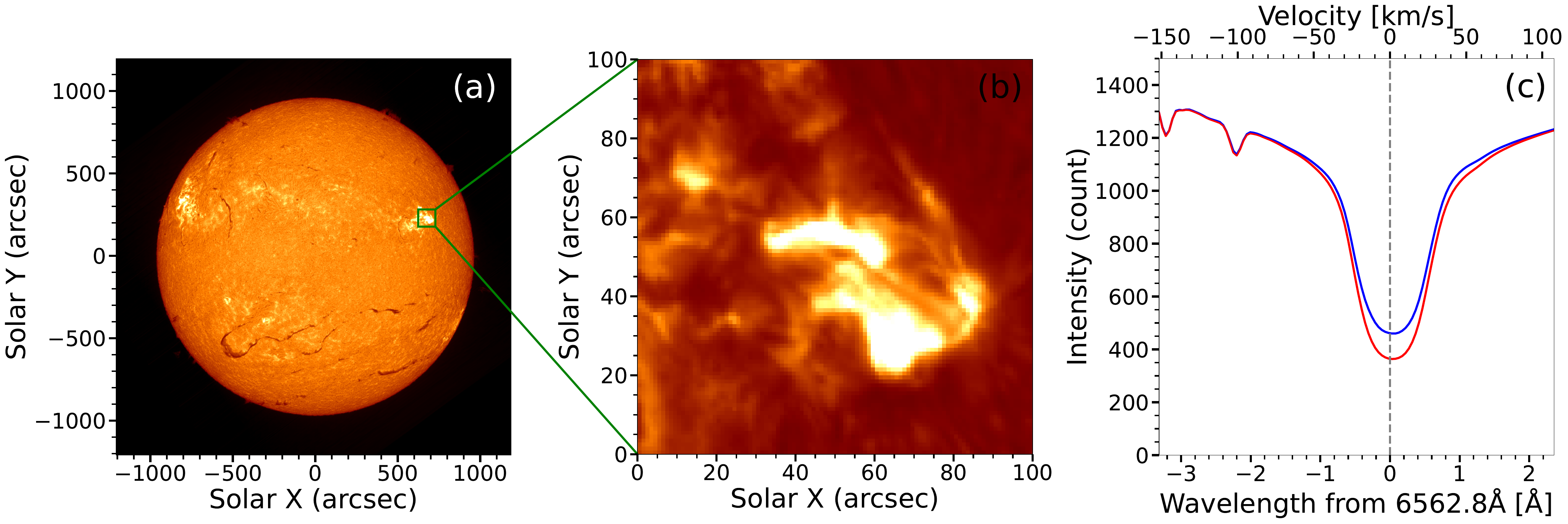}}
\caption{(a) CHASE $\mathrm{H}\alpha$ line center image of the full solar disk during the scan at $\text{20:25:19} \ \mathrm{UT} - \text{20:26:05} \ \mathrm{UT}$. 
The target region is indicated by the green box. 
(b) A zoom-in image of the TR. 
(c) Spatially averaged $\mathrm{H}\alpha$ profiles of the TR at $\text{20:25} \ \mathrm{UT}$ (the blue line) and at $\text{19:07} \ \mathrm{UT}$ (a pre-flare time, the red line), respectively. The grey dashed line marks the $\mathrm{H}\alpha$ line center.
\label{fig:2}}
\end{figure*}

First, we select a TR (the green region in Figure \ref{fig:2}a) and calculate the spatially averaged $\mathrm{H} \alpha$ profile of TR: 
\begin{equation}
    f_\mathrm{avg}(t, \lambda, \mathrm{TR}) = \frac{\int_{\mathrm{TR}}I(t, \lambda, x, y) \ \mathrm{d}x\mathrm{d}y}{N_\mathrm{pix}(\mathrm{TR})}, \label{eq1}
\end{equation}
where $I(t, \lambda, x, y)$ stands for the intensity at the observation time $t$, wavelength $\lambda$, and position $(x, y)$. 
$N_{\mathrm{pix}}(\mathrm{TR})$ is the number of pixels inside the TR. 
Second, we normalize $f_\mathrm{avg}(t, \lambda, \mathrm{TR})$ by the continuum level at a pre-flare time: 
\begin{equation}
    F_\mathrm{avg}(t, \lambda, \mathrm{TR}) = \frac{f_\mathrm{avg}(t, \lambda, \mathrm{TR})}{f_\mathrm{avg}(t, \lambda_\mathrm{cont}, \mathrm{TR})} \times f_\mathrm{avg}(t_{0}, \lambda_\mathrm{cont}, \mathrm{TR}), \label{eq2}
\end{equation}
where $t_{0}$ represents a pre-flare time and $\lambda_\mathrm{cont}$ represents a continuum wavelength. 
Note that $\lambda_\mathrm{cont}$ is selected from the blue wing of the $\mathrm{Fe} \ \mathrm{I}$ spectrum observed by CHASE. 
Thus, we obtain $F_\mathrm{avg}(t, \lambda, \mathrm{TR})$ (the blue line in Figure \ref{fig:2}c). 
This step is to suppress the influence coming from the temperature variation of HIS. 
Finally, we define the contrast profile $C(t, \lambda, \mathrm{TR})$ \citep{Hong} as 
\begin{equation}
    C(t, \lambda, \mathrm{TR}) = \frac{F_\mathrm{avg}(t, \lambda, \mathrm{TR}) - F_\mathrm{avg}(t_{0}, \lambda, \mathrm{TR})}{F_\mathrm{avg}(t_{0}, \lambda, \mathrm{TR})},  \label{eq3}
\end{equation}
where $F_\mathrm{avg}(t_{0}, \lambda, \mathrm{TR})$ is the spatially averaged profile of a pre-flare time $t_{0}$ (the red line in Figure \ref{fig:2}c). 

The contrast profile $C(t, \lambda, \mathrm{TR})$ indicates the fraction of the contribution in the spatial averaged profile coming from the TR, which is represented by $F_{avg}(t, \lambda, TR) - F_{avg}(t_{0}, \lambda, TR)$ divided by the spatial averaged profile of the same region at a pre-flare time $F_{avg}(t_{0}, \lambda, TR)$. 
If we assume that no other contributions are made by regions outside the TR, $C(t, \lambda, \mathrm{TR})$ can be interpreted as the contrast profile of the Sun when looking at it as a distant star, which certificates $C(t, \lambda, \mathrm{TR})$ as a Sun-as-a-star spectrum. 

\section{Result} \label{sec:Result}
\subsection{Influence of the Size of TR to the Sun-as-a-star Spectrum} \label{subsec:Result 1}
The ratio of the area of the solar flare to that of the TR declines as the TR enlarges. When TR is 
normalized to one pixel, the changes in the ratio can also be interpreted as the changes in the area of the solar flare. 

\begin{figure*}[ht!]
{\centering\includegraphics[width=1.0\textwidth]{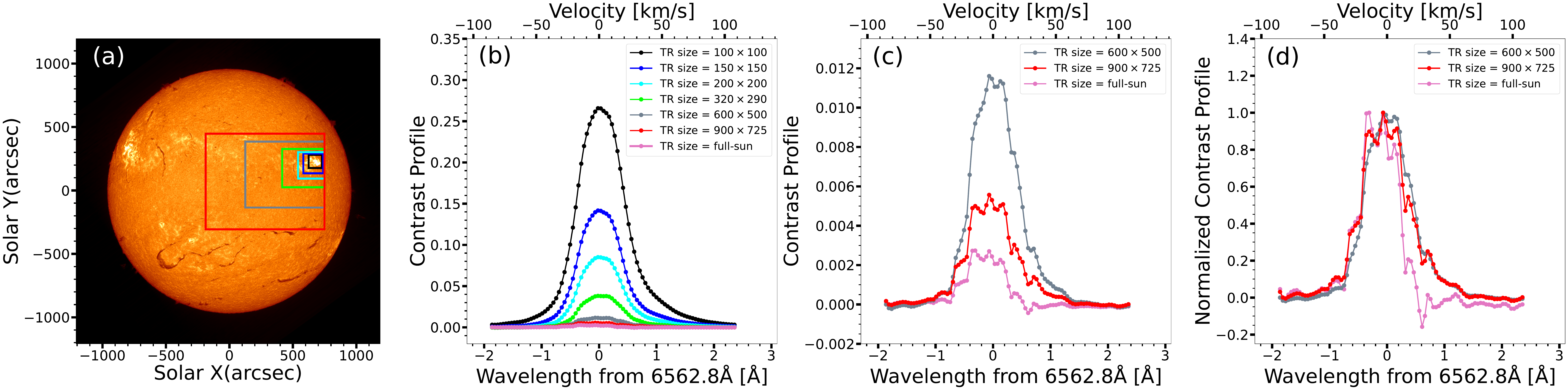}}
\caption{(a) $\mathrm{H}\alpha$ line center image of the full disk observed during $\text{20:25:19} \ \mathrm{UT} - \text{20:26:05} \ \mathrm{UT}$. 
TRs of different sizes are shown in different colors. 
(b) Contrast $\mathrm{H}\alpha$ profiles of the flare for different TRs. 
A contrast profile derived by integrating the full solar disk is also shown. 
(c) Contrast $\mathrm{H}\alpha$ profiles of the flare for the largest two boxes and the full disk  in panel (a). (d) Same as (c) but with their maximum values being normalized to $1$.
\label{fig:3}}
\end{figure*}

Here we study how the change in the ratio will affect the Sun-as-a-star profiles.
Figure \ref{fig:3} shows the variation of the Sun-as-a-star spectrum as the size of TR changes. 
As the TR gets larger, the peak of the corresponding Sun-as-a-star contrast profile gets weaker. 
It is clear that, in this case, for any TR with a size smaller than $900 \times 725\ \mathrm{pixels}$, emission enhancement at the $\mathrm{H}\alpha$ line center is discernible. 
At the same time, the features including line broadening and red wing enhancement can also be seen in the averaged spectra of those TRs. 
As shown in Figure \ref{fig:3}c, for the TR with a size of $900 \times 725\ \mathrm{pixels}$, emission enhancement in the $\mathrm{H}\alpha$ line center and line broadening can still be observed. Figure \ref{fig:3}d yields a clearer look for the three Sun-as-a-star spectra in Figure \ref{fig:3}c as their maximum values are normalized to 1, in which the Sun-as-a-star profile within TR of $900 \times 725 \ \mathrm{pixels}$ presents as a slightly red asymmetric one.

The full-disk profile, on the other hand, is an emission line with a red absorption characteristic (Figure \ref{fig:3}d). This could be caused by two reasons. (\romannumeral1) Due to the oscillation, a quiescent filament moved downward and thus showed a red-shifted absorption around 20:25 UT \citep{2023ApJDai}. This could influence the full-disk Sun-as-a-star profiles of the flare. (\romannumeral2) Our selected pre-flare data (the last scanning of CHASE during the previous orbit to the orbit that captured the X1.0 flare) inevitably includes a C8.1 flare. It took place in NOAA AR 13112 from $\text{18:49} \ \mathrm{UT}$ to $\text{19:25} \ \mathrm{UT}$. This solar flare is again not included in any of our TRs, so it has an impact only on the full-disk Sun-as-a-star profile. This flare may lead to an enhancement in the red wing of the pre-flare full-disk profile, thus a depletion in the red wing of the full-disk Sun-as-a-star contrast profile. This indicates that Sun-as-a-star profiles derived from full-disk integration may not necessarily reveal the properties of the solar activity of interest. Note that any TRs in our analyses do not include the filament or the C8.1 flare except for the full disk integration, which means that our main results regarding the Sun-as-a-star properties of the X1.0 solar flare should not be changed by these two possibilities.

Given the fact that the CHASE observes the Sun with high spectral resolution, small random noise may also be captured by HIS and reflected in the Sun-as-a-star $\mathrm{H}\alpha$ profile. Also, it is sometimes inevitable that the TR contains extra small-scale solar activities (i.e. floccules or small-scale filament eruptions). All these above will consequently lead to a Sun-as-a-star $\mathrm{H}\alpha$ profile with fluctuations. We can view these fluctuations as background noise in the spectrum brought by observation errors and other solar events. These fluctuations are more obvious in the contrast profiles with relatively low intensity. For instance, the integral result of a TR of size $900 \times 725\ \mathrm{pixels}$ (Figure \ref{fig:3}c) shows such a fluctuation, particularly around the $\mathrm{H}\alpha$ line center. However, for contrast profiles with high intensity (derived from small TRs), the influence of random noise may not be significant, but it may lead to an unsmooth Sun-as-a-star spectrum. It should be noted that, even though these fluctuations in the spectrum may reflect some solar events, they can be of influence when we want to investigate the major spectroscopic characteristics of the Sun-as-a-star profiles.
In order to quantitatively diagnose spectral properties, like line broadening and red asymmetry, we propose to use discrete Fourier transform (DFT) as a method to reduce the noise while preserving the main features of the spectra. 

DFT is a mature denoising algorithm that has been applied in areas involving high spectral resolution data, like High Definition (HD) FT-IR and Quantum Cascade Laser (QCL) Microscopes imaging analysis for a long time \citep{RN48, RN49, RN50}. It deduces the original data into sinusoids, each with a different wavenumber, while the sum of them makes up the original signal. This operation is often referred to the transformation from the wavelength domain to the wavenumber domain. Random noise in the Sun-as-a-star profiles generally corresponds to the high wavenumber part in the wavenumber domain. After using window operations to suppress the high wavenumber part, we apply the reverse DFT to the wavenumber domain so as to acquire a profile with diminished noise. Among all denoising algorithms, such as wavelets transformation, Principal Component Analysis (PCA) and Minimum Noise Fraction (MNF), DFT has been proven to be an efficient way to denoise signals with low signal distortion and short computational time \citep{RN51}.
Here we apply DFT with a super-Gaussian window to the contrast profiles as proposed by \citet{RN47}. 
This method will reduce the noise in our profiles without broadening the peak, which means that it ensures our diagnosis regarding the line broadening and red asymmetry properties of the denoised signal to be accurate \citep{RN47}. 

After we acquire the denoised spectrum, a three-component fitting, which has been often applied in stellar flare observations \citep{RN46}, is employed here to identify the line broadening and red asymmetry of the spectrum. 
The three components consist of a Voigt function, a Gauss function, and a constant. 
The Voigt function is to fit the emission at the $\mathrm{H}\alpha$ line center, the Gauss function is to fit the red wing enhancement, and the constant is to fit the background emission. 
The mathematical expression of the fitting function can be written as 
\begin{equation}
    F(\lambda) = \frac{\mathrm{Re}\left[ \mathrm{wofz}\left( \frac{\lambda - \lambda_\mathrm{cen} + \mathrm{i}\gamma}{\sqrt{2}\sigma} \right) \right]}{\sqrt{2\pi}\sigma}I_{V} + 
    I_{G} \mathrm{e}^{-\left( \frac{\lambda - \lambda_{G}}{s} \right)^{2}} + C, \label{eq4} 
\end{equation}
where $\lambda_\mathrm{cen}$ and $I_{V}$ are the line center and total flux of the Voigt profile, respectively. $\mathrm{Re}$ means taking the real part of the complex number, and $\mathrm{wofz}$ is the Faddeeva function. Note that Voigt function can be taken as the convolution of Lorentzian and Gaussian functions, thus in equation (\ref{eq4}) $\gamma$ represents the half width at half maximum of the Lorentzian function, and $\sigma$ is the standard deviation of the Gaussian function. For the Gaussian component in equation (\ref{eq4}), $I_{G}$ is the maximum intensity, $\lambda_{G}$ is the centroid wavelength and $s$ is the standard deviation of the Gaussian component. $C$ is the constant component. 

\begin{figure}[htbp]
\centering
{\includegraphics[width=0.9\linewidth]{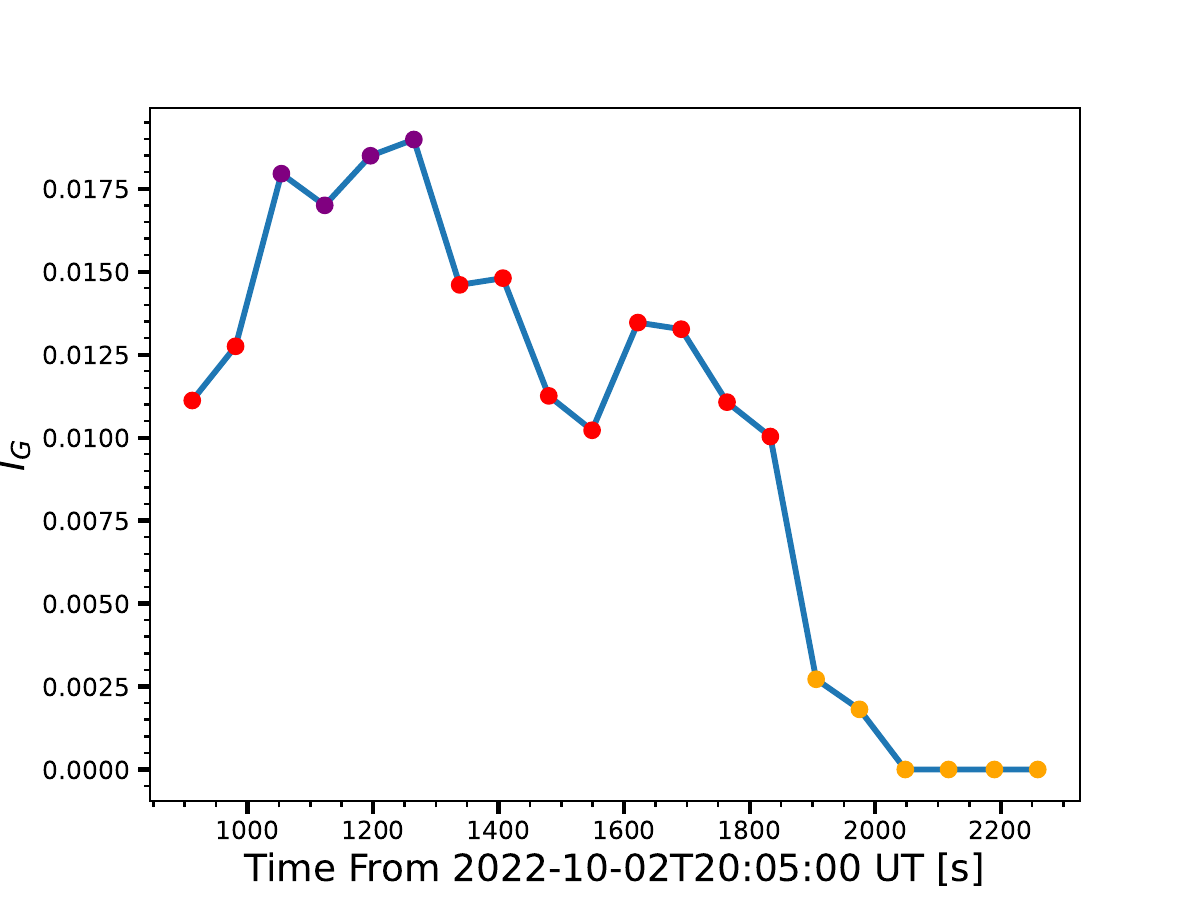}}
\caption{$I_{G}$ evolution with time derived from the three-component fitting results of the TR with size $100 \times 100 \ \mathrm{pixel}$. The purple, red, and yellow dots indicate the high, medium, and low intensity $I_{G}$, respectively.}  
\label{fig:4}
\end{figure}

It is found that for each TR, the Gaussian component's intensity has a decreasing trend against time (Figure \ref{fig:4}). After the end time of the flare, all spectra can be well-fitted without the Gaussian component. For those spectra, if forced to be fitted by the three-component function, the intensity of the Gaussian component will yield a very small value. In this case, in order to constrain errors, a new fitting without the Gaussian component is adopted. In order to effectively tell apart which profile needs to be fitted with a Gaussian component and which not, we use the K-means algorithm \citep{ostrovsky2013effectiveness} to separate the three-component fitting results into different classes based on the intensity of the Gaussian component. In our case, we suggest to set $\mathrm{K}$ as $3$ to classify Gaussian components with high, medium and low intensities (Figure \ref{fig:4}). Those profiles that have a low Gaussian component intensity will be refitted with the function excluding the Gaussian function.

\begin{figure*}[ht!]
{\centering\includegraphics[width=1.0\textwidth]{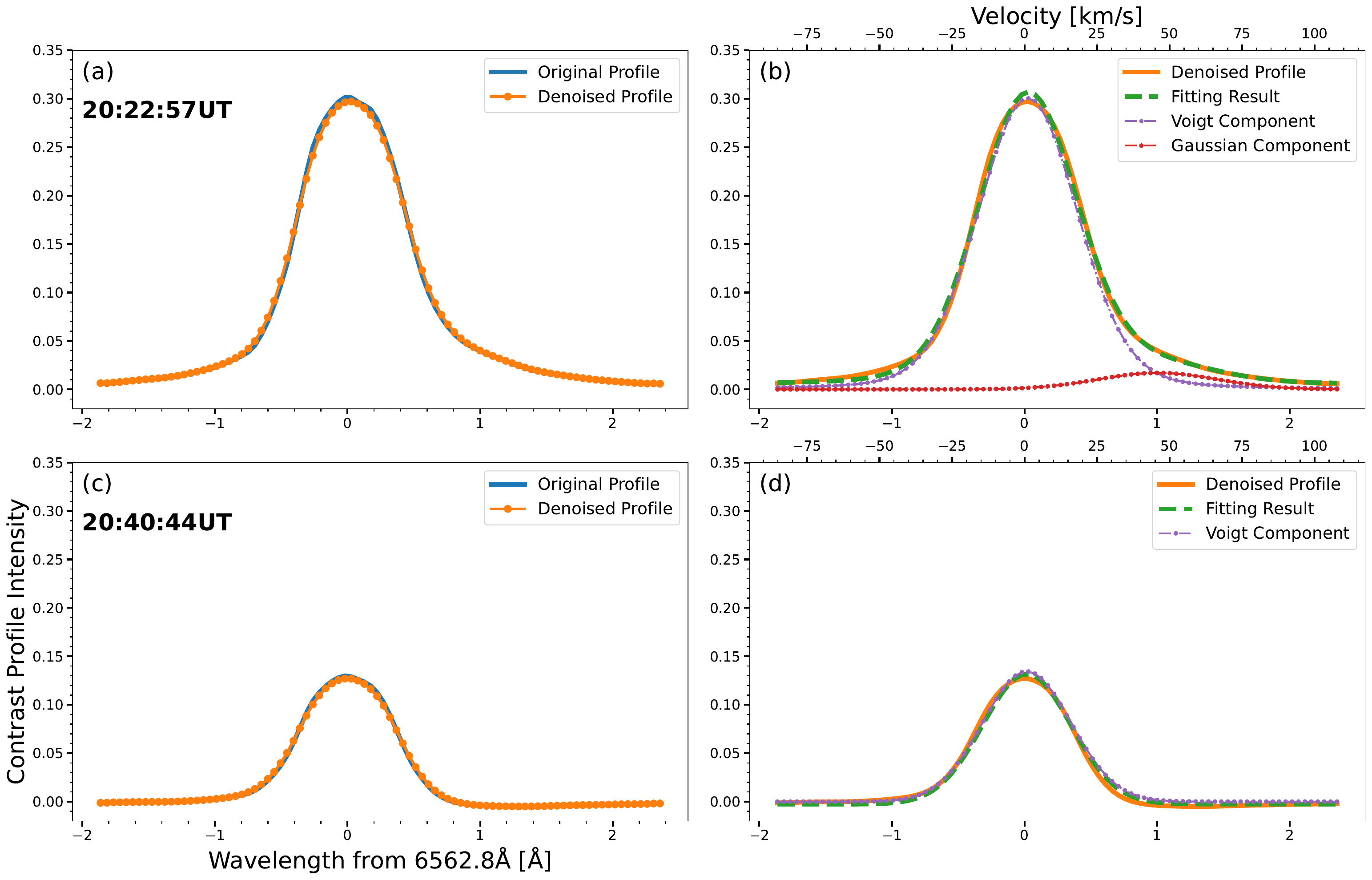}}
\caption{The fitting process of the Sun-as-a-star $\mathrm{H}\alpha$ contrast profile of a TR. 
(a) The Sun-as-a-star $\mathrm{H}\alpha$ contrast profile calculated based on the scan started at $\text{20:22:57} \ \mathrm{UT}$ (the blue line) and its denoised result (the orange line). 
(b) The fitting result of the denoised $\mathrm{H}\alpha$ profile. 
The orange line represents the denoiesd spectrum. 
The green dashed line is the fitting result which consists of a Voigt component, a Gaussian component and a constant. 
The purple dotted line is the Voigt component and the red dotted line is the Gaussian component. 
(c) - (d) The denoised and fitting result for the Sun-as-a-star spectrum based on the scan started at $\text{20:40:40} \ \mathrm{UT}$. 
Note that this fitting is done without a Gaussian component (the green dashed line). 
\label{fig:5}}
\end{figure*}

The total procedure of denoising and fitting can be expressed as follows: 
First, we denoise the Sun-as-a-star spectrum using DFT with a super-Gaussian window and acquire the denoised spectrum prepared for the fitting (Figure \ref{fig:4}a, c). 
Second, we determine the center wavelength of the Voigt function. 
We use the center-of-gravity method \citep{RN45} to calculate the $\mathrm{H}\alpha$ line center of the TR at $t_{0}$ (a pre-flare time) and take it as the line center ($\lambda_\mathrm{cen}$) of the Voigt function. 
Third, we fit the summation of the Voigt function, the Gauss function and the constant to the denoised spectrum of a certain TR size at all different times to determine all free parameters. 
In the end, we check the $I_{G}$ in the fitting result, and use the K-means method described above to separate the fitting results. If the $I_{G}$ falls into the high or medium intensity criteria, the fitting result is accepted (Figure \ref{fig:5}b), otherwise we certify that this Gaussian component is too weak to be considered in the fitting. 
In this case, a new fitting without considering the Gaussian function is employed (Figure \ref{fig:5}d). 

\begin{figure*}[ht!]
{\centering\includegraphics[width=1.0\textwidth]{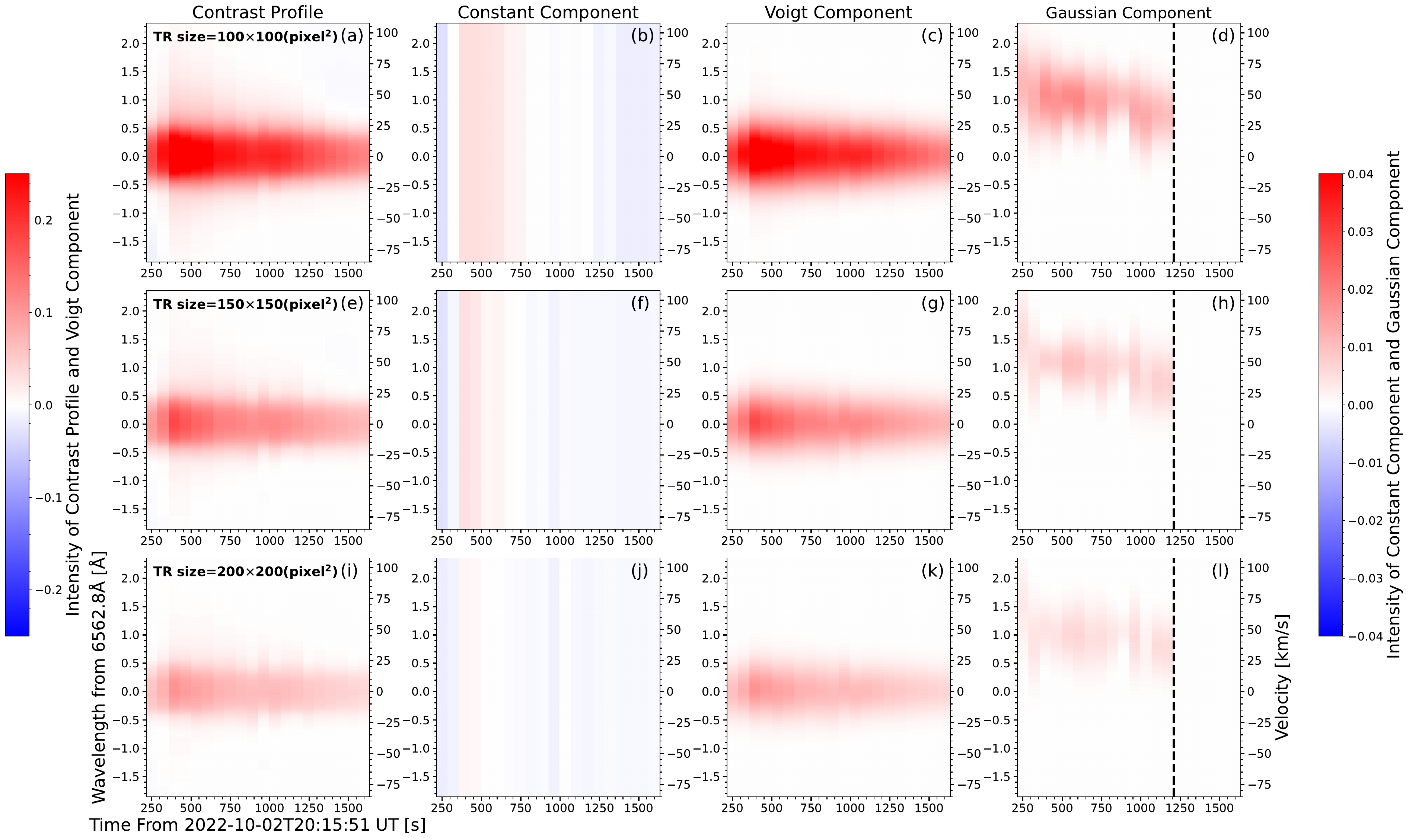}}
\caption{Temporal evolution of $\mathrm{H}\alpha$ contrast spectra (a) and corresponding fitting results (b-d) for TR size of $100\times100\ \mathrm{pixels}$ (the black square in Figure \ref{fig:3}a). 
(b), (c) and (d) refer to the constant component, Voigt component and Gaussian component, respectively. 
(e-h) and (i-l) are the same as (a-d) but for different TR sizes. The black dashed lines in (d), (h) and (l) mark $1209 \ \mathrm{s}$ from $\text{20:15:51} \ \mathrm{UT}$.
\label{fig:6}}
\end{figure*}

Figure \ref{fig:6} shows the Sun-as-a-star dynamic spectra of TRs with three different sizes (each corresponding to the black, blue, and cyan square in Figure \ref{fig:3}). 
For all of the Sun-as-a-star dynamic spectra, clear emission enhancement could be seen at the $\mathrm{H}\alpha$ line center (Figure \ref{fig:6}a, e, and i). 
Dynamic spectra of the constant component, the Voigt component, and the Gaussian component reveal more details of the flare. 
The dynamic spectra of the constant component have an obvious feature, which indicates that the flare is a white light one. 
As the TR gets larger, the emission intensity weakens (Figure \ref{fig:6}b, f, and j). 
The dynamic spectra of the Voigt component give a clear picture of the long-lasting emission near the $\mathrm{H}\alpha$ line center. 
The emission is preserved even after the end time of the flare as defined from GOES soft X-ray flux curve. 
The dynamic spectra of the Voigt component also show a decaying trend in both the peak intensity and the line width as the TR size increases (Figure \ref{fig:5}c, g, and k). 
The Gaussian component shows the redshift velocity of the flare to be around $50 \ \mathrm{km \ s^{-1}}$ and it shows a decaying trend in the early phase against time, which is especially obvious in Figure \ref{fig:6}d. 
The intensity of the Gaussian component is weaker than that of the Voigt component, and it also decreases as the TR gets larger (Figure \ref{fig:5}d, \ref{fig:6}h, and \ref{fig:6}l). 
Note that, after $1209 \ \mathrm{s}$ since $\text{20:15:51} \ \mathrm{UT}$ in all three TRs, the Sun-as-a-star profiles can be well explained without a Gaussian component (i.e. the dynamic spectra of the Gaussian component show no signal after $1209 \ \mathrm{s}$), which means that the Gaussian component of the Sun-as-a-star profile decays to zero earlier than the Voigt component. 
This infers that, during the late phase of the flare, only emission enhancement in the $\mathrm{H}\alpha$ line center in the Sun-as-a-star profile can be observed.

\subsection{Variation of SNR and Redshift Velocity with TR size} \label{subsec:Result 2}
As discussed in Section \ref{subsec:Result 1}, an apparent trend can be seen in the dynamic spectra that as the TR gets larger, the peak intensity of the contrast profile weakens. 
At the same time, the redshift velocity has a typical value of around $50 \ \mathrm{km \ s^{-1}}$. 
However, how the size of the TR influences the signal to noise ratio (SNR) and redshift velocity of a Sun-as-a-star spectrum in detail still remains unclear. 
Here we study the variation of SNR and redshift velocity with the size of TR. 
We choose 11 TRs with sizes ranging from $100\times100 \ \mathrm{pixels}$ to $200\times200 \ \mathrm{pixels}$ in a interval of $10 \ \mathrm{pixels}$ in side length (the green regions in Figure \ref{fig:6}). 
We define $\mathrm{SNR}$ as 
\begin{equation}
    SNR = 10\times \log_{10}\left( \frac{\sum_{i} (C_\mathrm{dn}(\lambda_{i}))^{2}}{\sum_{i} (C(\lambda_{i}))^{2} - \sum_{i} (C_\mathrm{dn}(\lambda_{i}))^{2}} \right), \label{eq5}
\end{equation}
where $C(\lambda)$ represents the contrast profile and $C_\mathrm{dn}(\lambda)$ represents the denoised contrast profile. 
The value of SNR should be taken as an indication of the quality of the Sun-as-a-star profile, as it is the proportion of the signal's intensity and noise's intensity in the profile. 
Thus, the profiles with larger SNR should give more trustable spectral characteristics of the flare than those with smaller SNR. 

\begin{figure*}[ht!]
{\centering\includegraphics[width=1.0\textwidth]{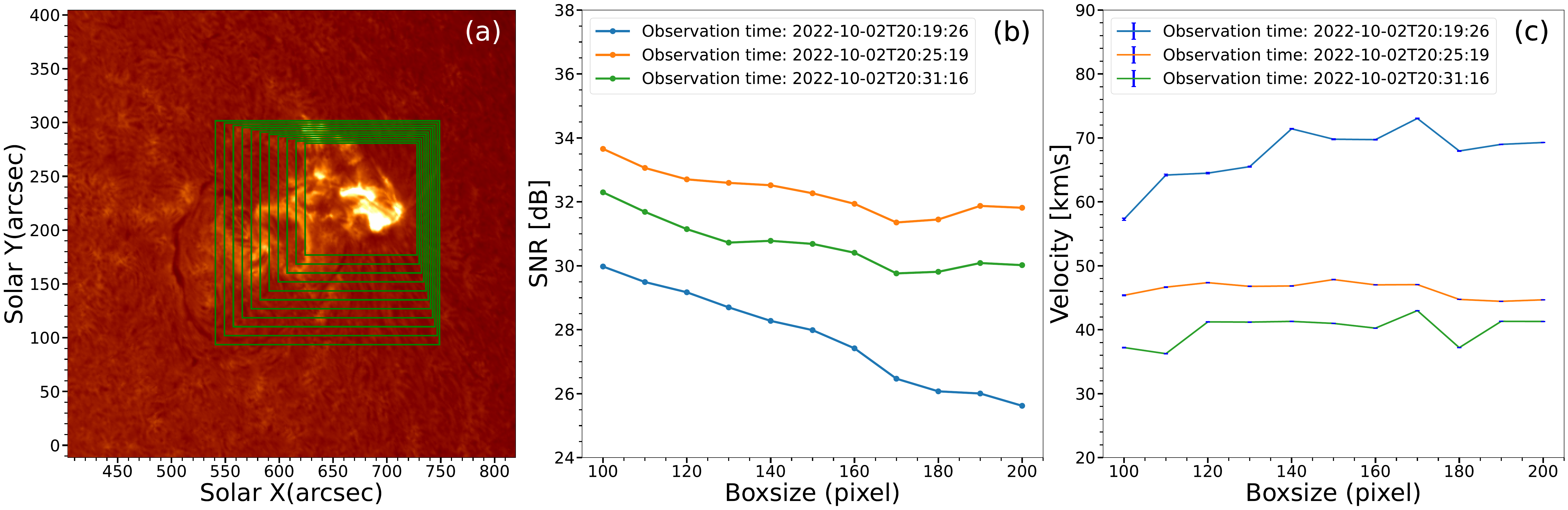}}
\caption{(a) Part of field-of-view of CHASE $\mathrm{H}\alpha$ line center image of the flare, which was observed during the scan at $\text{20:25:19} \ \mathrm{UT} - \text{20:26:05} \ \mathrm{UT}$. 
The different boxes with different sizes show TRs. 
(b) Evolution of the SNR with the size of TR at three different time. 
(c) Evolution of the redshift velocity with the size of TR at three different time. 
The blue bars are the errors from the fitting. 
\label{fig:7}}
\end{figure*}

Figure \ref{fig:7}b shows the variation of SNR with the TR size at three different time. 
A steady decrease can be seen in SNR as TR gets larger. 
Meanwhile, all three lines in Figure \ref{fig:7}a reveal a similar downtrend, which shows that the size of TR influences SNR of the Sun-as-a-star profile in a unified way. 
The decrease in SNR as the size of the TR gets larger indicates that the signal becomes harder to be identified from the background noise. 

The redshift velocity can be easily derived from the fitting results as 
\begin{equation}
    v_\mathrm{red} = \frac{\lambda_\mathrm{G} - \lambda_\mathrm{cen}}{\lambda_\mathrm{cen}}c, \label{eq6}
\end{equation}
where $\lambda_\mathrm{G}$ stands for the line center of the Gaussian component, and $\lambda_\mathrm{cen}$ stands for the line center of the mean $\mathrm{H}\alpha$ profile of the TR at a pre-flare time $t_{0}$. 

Figure \ref{fig:7}c shows the variation of redshift velocity with the size of TR at three different time. 
It can be seen that, as the TR increases, the redshift velocity remains a steady level though with very little fluctuations. 
The consistency of redshift velocity predicates that some dynamics features of the flare maintain despite changes in the size of TRs. 
Note that the small error bars can be taken as a proof of the accuracy of the fitting. 

\subsection{The Upper Limit in the Size of TR with Red Asymmetry} \label{subsec:Result 3}
From Section \ref{subsec:Result 1}, we find that the intensity of the Gaussian component decays to zero earlier than that of the Voigt component. 
We can also see that as TR gets larger, the intensity of the Gaussian component weakens. Given that the Gaussian component measures the red asymmetry in the Sun-as-a-star profile, we wonder if there is an upper limit in the size of TR at which the Sun-as-a-star profile can show red asymmetry at the peak time of the flare. 
This question can swiftly transform into the question of whether there is a lower limit in spatial resolution, at which the $\mathrm{H}\alpha$ profile of a stellar flare does not show red asymmetry even at its peak time. 

\begin{figure*}[ht!]
{\centering\includegraphics[width=1.0\textwidth]{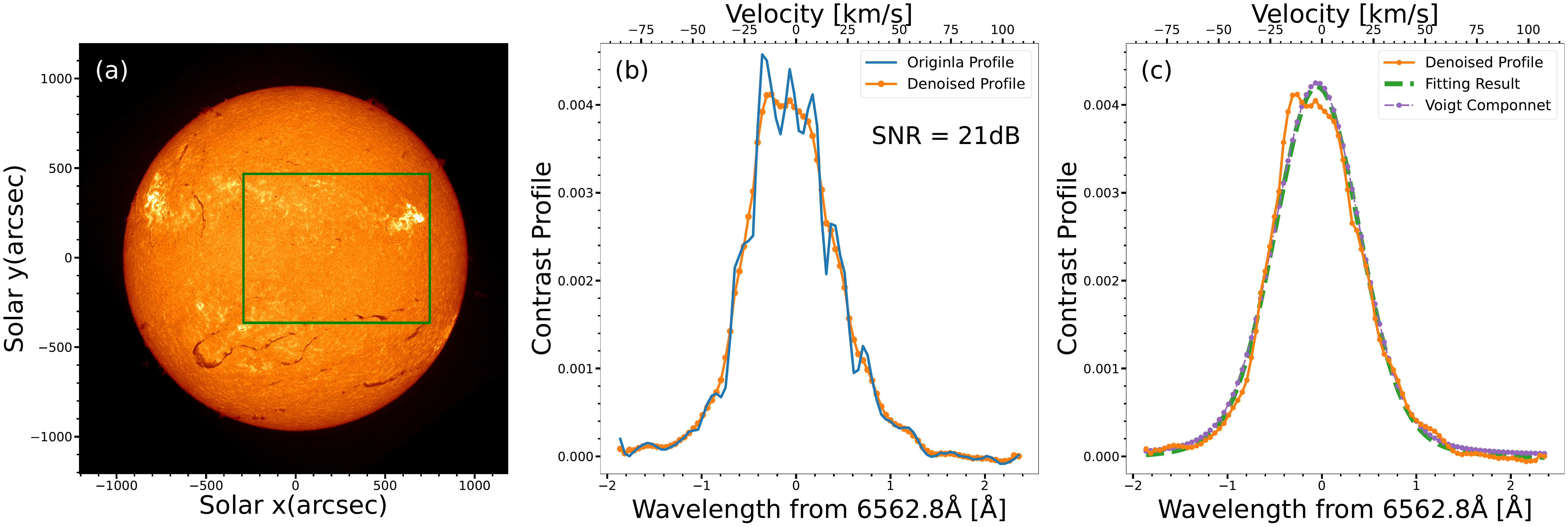}}
\caption{(a) An $\mathrm{H}\alpha$ line center image taken during the scan at $\text{20:25:19} \ \mathrm{UT} - \text{20:26:05} \ \mathrm{UT}$. 
The TR is shown by the box that contains $1000 \times 800 \ \mathrm{pixels}$. 
(b) Sun-as-a-star $\mathrm{H}\alpha$ contrast profile of the flare for the TR (the blue line) and its denoised result (the orange line). 
(c) The fitting result of the $\mathrm{H}\alpha$ contrast profile. 
\label{fig:8}}
\end{figure*}

In order to find the lower limit of the size of TR that does not have red asymmetry, we utilize the HIS image taken around the peak time of the GOES SXR light curve. 
We choose TR around the flare site and gradually increase the size of the TR. 
Then, we calculate the Sun-as-a-star contrast profile based on the TR. 
The denoised algorithm and the three-component fitting are then performed on the Sun-as-a-star profile as explained in Section \ref{subsec:Result 1}. 
After back-and-forth tries, we consider that the TR with a size of $1000 \times 800 \ \mathrm{pixels}$ ($1 \ \mathrm{pixel} = 1.04\times1.04 \ \mathrm{arcsec}^{2}$) to be the smallest TR at which the Sun-as-a-star profile shows no red asymmetry characteristic (Figure \ref{fig:8}a). 
The profile only shows an emission feature at the $\mathrm{H}\alpha$ line center (Figure \ref{fig:8}b). 
The SNR of this specific profile is determined as $21 \ \mathrm{dB}$. 
The fitting result further illustrates that the features of the Sun-as-a-star profile can be well explained without the existence of the Gaussian component in the red wing (Figure \ref{fig:8}c). 

This result confirms the existence of TRs whose corresponding Sun-as-a-star profiles at the flare time can only be observed as $\mathrm{H}\alpha$ emission profiles without any diagnosable red wing enhancements. 
It is worthy of noting that, even if the red wing enhancement becomes very small as the TR enlarges, we may still be able to observe it if there were no fluctuations in the spectrum. 
However, fluctuations do exist, which means that as the size of TR expands, the red wing enhancement may become indistinguishable from background noise and go unnoticed. 
In simpler terms, in this specific case we analyzed, the minimum SNR required to detect red asymmetry in the $\mathrm{H}\alpha$ line, which indicates chromospheric dynamics during a flare, is $21 \ \mathrm{dB}$. 
Any sun-as-a-star profile with an SNR value lower than this threshold cannot reveal the full properties of the flare. It should be emphasized that, this upper limit in the size of the TR with red asymmetry and its corresponding SNR is derived based on CHASE's observation and this specific solar flare.

\subsection{Time Evolution of the \texorpdfstring{$\mathrm{H}$}{}\texorpdfstring{$\alpha$}{} EW and Redshift Velocity} \label{subsec:Result 4}
We also study the time evolution of the $\mathrm{H}\alpha$ equivalent width (EW) and redshift velocity of the Sun-as-a-star spectrum. 
We calculate the EW of the Sun-as-a-star $\mathrm{H}\alpha$ profile as $\int C(t, \lambda, \mathrm{TR}) \ \mathrm{d}\lambda$. 
We choose three TRs with different sizes: $100\times100 \ \mathrm{pixels}$ (the region inside the black square in Figure \ref{fig:3}), $150\times150 \ \mathrm{pixels}$ (the region inside the blue square in Figure \ref{fig:3}), and $200\times200 \ \mathrm{pixels}$ (the region inside the cyan square in Figure \ref{fig:3}), and examine their Sun-as-a-star profiles at each different time. 
Note that as mentioned in Section \ref{subsec:Result 1}, for Sun-as-a-star profiles calculated based on scans after $1209 \ \mathrm{s}$ since $\text{20:15:51} \ \mathrm{UT}$, the intensity of the Gaussian component is zero. 
These moments are not taken into account as we discuss the time evolution of the redshift velocity. 

\begin{figure*}[ht!]
{\centering\includegraphics[width=1.0\textwidth]{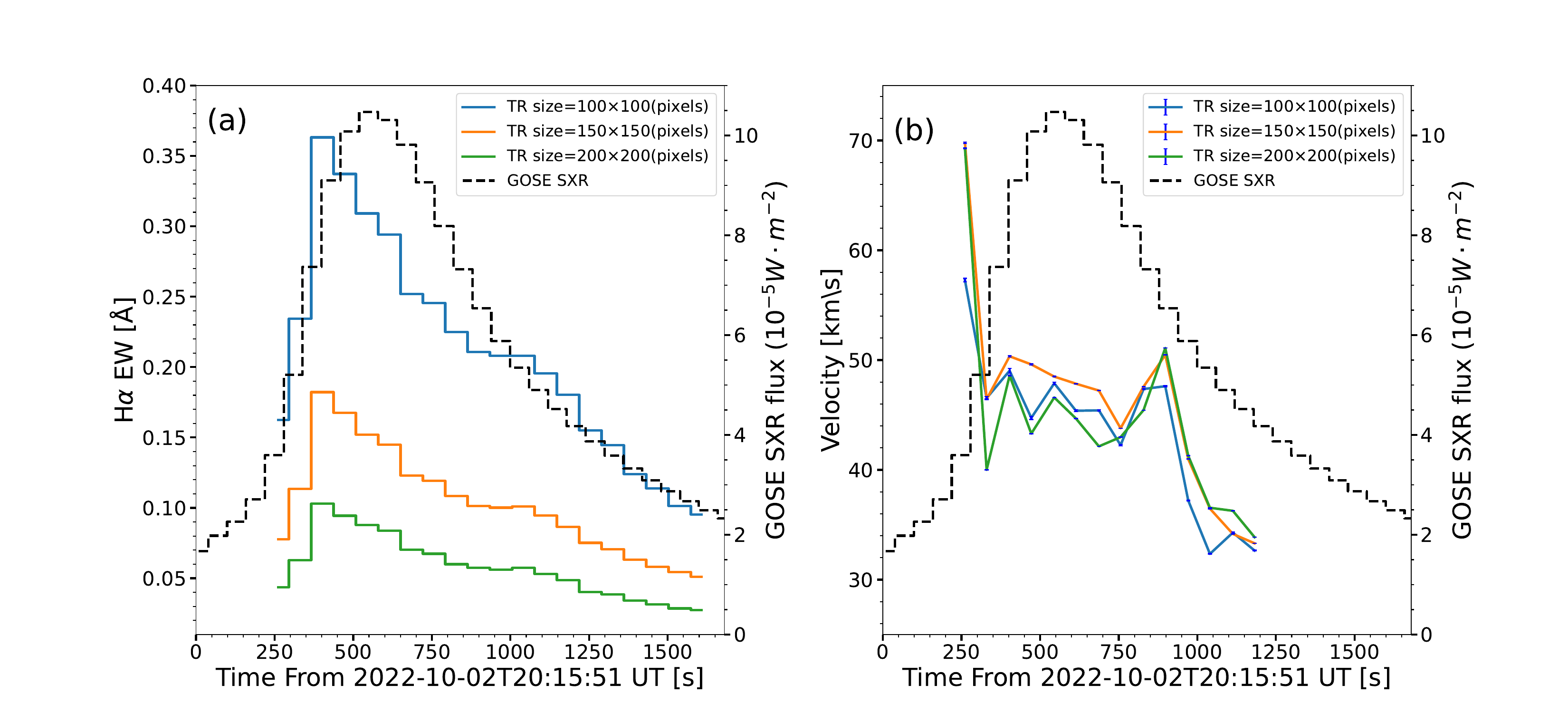}}
\caption{(a) Temporal evolution of the Sun-as-a-star $\mathrm{H}\alpha$ EW of the flare for three different TRs. The GOES 1--8 {\AA} SXR flux is plotted as the black dashed line.
(b) Temporal evolution of the redshift velocity for three different TRs. 
The blue bar represents the fitting errors. 
\label{fig:9}}
\end{figure*}

From Figure \ref{fig:9}a, we can see a similar rise and fall trend in the time evolution of the $\mathrm{H}\alpha$ EW of the Sun-as-a-star profiles in all three TRs. 
The $\mathrm{H}\alpha$ EW shows a quick rise to peak time followed by a gradual decrease, similar to the GOES SXR. 
Their peak times show a very slight difference. 

Figure \ref{fig:9}b shows the time evolution of the redshift velocity of the three different TRs. 
All three lines overlap each other, which confirms our conclusion in Section \ref{subsec:Result 2} that the change in TR size will not cause a significant influence on the detection and estimation of redshift velocity. 
The redshift velocity obviously decreases over time. Although it's difficult to determine the exact time of the peak redshift velocity with our data, we can conclude that the peak time of the redshift velocity should occur before the peak time of $\mathrm{H}\alpha$ EW, and therefore before the peak time of GOES SXR. 

\subsection{Time Average of \texorpdfstring{$\mathrm{H}$}{}\texorpdfstring{$\alpha$}{} EW and Redshift Velocity} \label{subsec:Result 5}
Here we consider the properties of temporally averaged $\mathrm{H}\alpha$ spectra. 
Our main purpose is to study how the Sun-as-a-star contrast profile changes in $\mathrm{H}\alpha$ EW and its Gaussian component after it is further averaged in time. 
We first define the time-averaged profile at the pixel level as 
\begin{equation}
    I_\mathrm{avg}(\lambda, x, y) = \frac{\int_{T_{0}}^{T_{1}} I(t, \lambda, x, y) \ \mathrm{d}t}{T_{1} - T_{0}} \label{eq7}
\end{equation}
where $I(t, \lambda, x, y)$ is the $\mathrm{H}\alpha$ profile at position $(x, y)$ and time $t$, and $I_\mathrm{avg}(\lambda, x, y)$ represents the time average profile between $T_{0}$ and $T_{1}$ at pixel level. 
After acquiring the time-averaged profile at pixel level, we can use it to substitute the $I(t, \lambda, x, y)$ in equation (\ref{eq1}), thus obtaining the time-averaged Sun-as-a-star $\mathrm{H}\alpha$ profile by following the steps mentioned in Section \ref{subsec:Method 2}. 
Once we have the time-averaged Sun-as-a-star $\mathrm{H}\alpha$ profile, we immediately apply the denoising and fitting method introduced in Section \ref{subsec:Result 1} and then derive the Gaussian component of the profile. 
The EW of the Gaussian component is defined similarly to that of the $\mathrm{H}\alpha$ profile. 
It equals to $\int I_{G} \mathrm{e}^{-\left( \frac{\lambda - \lambda_{G}}{s} \right)^{2}} \ \mathrm{d}\lambda$, where $I_{G}$, $\lambda_{G}$, and $s$ are determined by the fitting result. 
In this study, we adopt HIS $\mathrm{H}\alpha$ image taken by CHASE between $\text{20:19:26} \ \mathrm{UT}$ and $\text{20:41:53} \ \mathrm{UT}$ and analyze their time-averaged Sun-as-a-star profiles. 
The TR is set with a size of $100\times100 \ \mathrm{pixels}$ covering the flare (the region inside the black square in Figure \ref{fig:3}). 

\begin{figure*}[ht!]
{\centering\includegraphics[width=1.0\textwidth]{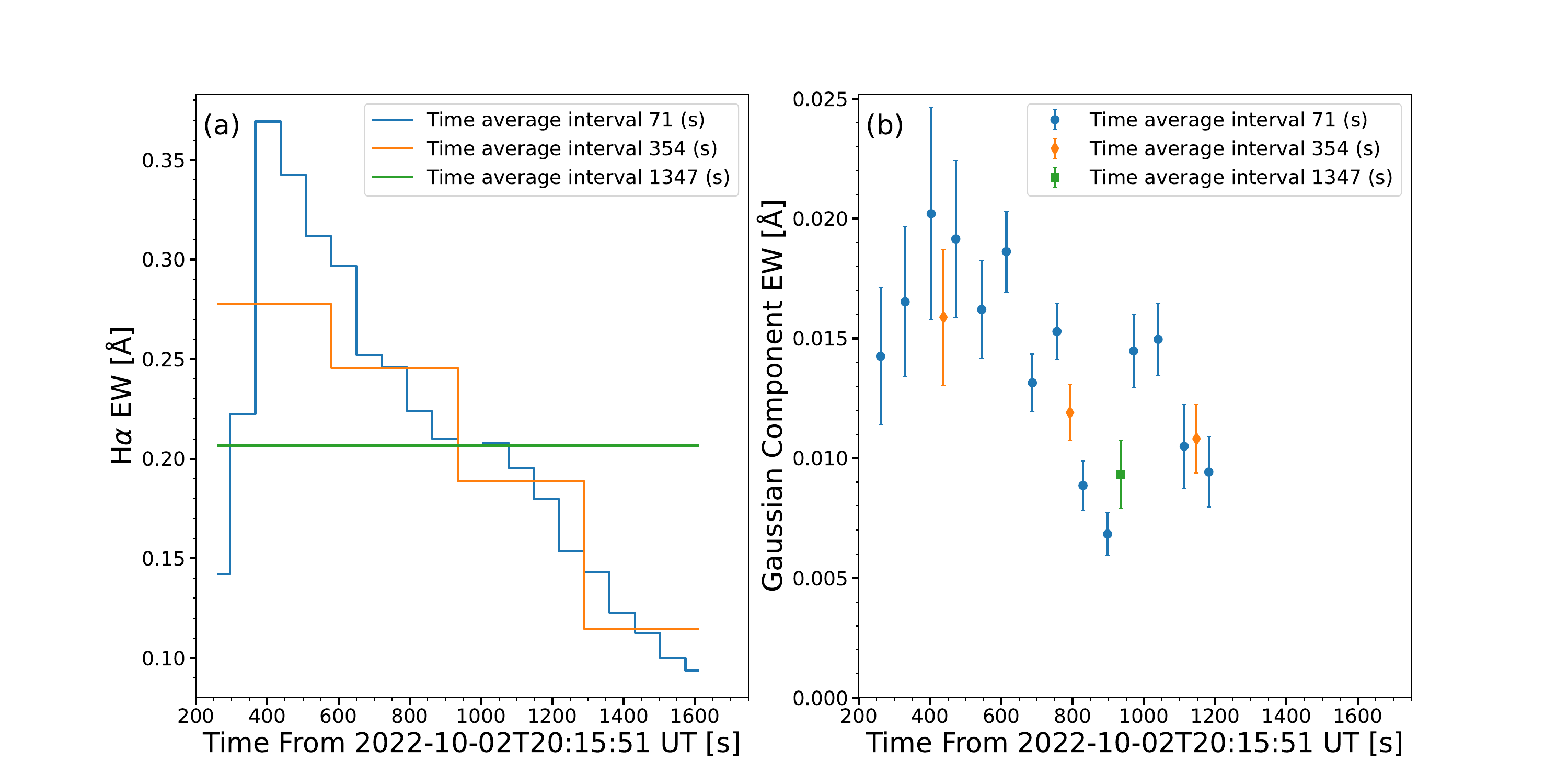}}
\caption{Variations of the $\mathrm{H}\alpha$ EW (a) and Gaussian component EW (b) as the integral time increases. The errors as shown by the vertical bars are from the fitting. 
\label{fig:10}}
\end{figure*}

Figure \ref{fig:10}a shows the time evolution of $\mathrm{H}\alpha$ EW corresponding to different time average intervals. 
The results without averaging in time are drawn as references. 
In contrast, the result corresponding to the $354 \ \mathrm{s}$ interval only exhibits a decaying trend. 
Meanwhile, the peak value of the EW is obviously lower than (only about $77\%$ of) that for without averaging. 
We also consider the case of a $1347 \ \mathrm{s}$ time average interval, which represents the time average of all Sun-as-a-star $\mathrm{H}\alpha$ profiles observed by CHASE between $\text{20:19:26} \ \mathrm{UT}$ and $\text{20:41:53} \ \mathrm{UT}$. 
This represents the extreme condition that the solar flare is observed without time resolution. 
It can be seen that, even in this situation, the $\mathrm{H}\alpha$ EW remains above zero, which at least indicates an emission at the $\mathrm{H}\alpha$ line center. 
It is also worth noticing that, under this circumstance (i.e. $1347 \ \mathrm{s}$ time average interval), the $\mathrm{H}\alpha$ EW is between the maximal and minimal $\mathrm{H}\alpha$ EWs of the result coming from the $71 \ \mathrm{s}$ time average interval. 

Figure \ref{fig:10}b shows the temporal evolution of the Gaussian component EWs derived from the Sun-as-a-star profiles with different time-average intervals. 
Note that the time-average profiles that can be fitted without a Gaussian component are not taken into consideration. 
For $71 \ \mathrm{s}$ time-average interval, we can see that the Gaussian component EW has an overall decaying trend with fluctuations. 
Meanwhile, the Gaussian component EW with $354 \ \mathrm{s}$ interval only shows a decaying trend, with the maximum value lower than that with $71 \ \mathrm{s}$ interval. 
The Gaussian component EW with $1347 \ \mathrm{s}$ interval is also above zero. 
Combined with the $\mathrm{H}\alpha$ EW of the same time average interval, we may conclude that, for this case, even without time resolution, the Sun-as-a-star contrast profile still exhibits an emission feature in $\mathrm{H}\alpha$ line center as well as a red asymmetry. 
However, even though we can observe both the line center emission and red asymmetry, these two features suggest a significant decrease in magnitude compared with the time-resolved results. 

\section{Summary and Discussion} \label{sec:Summary}

We analyze the Sun-as-a-star $\mathrm{H}\alpha$ spectra of an X1.0 solar flare taking advantage of CHASE/HIS data and reveal two major characteristics: long-lasting emission enhancement in the $\mathrm{H}\alpha$ line center with line broadening, and red asymmetry, which is consistent with previous studies \citep{RN62,RN63}. 
Additionally, we use the three-component fitting method to study the red asymmetry of the $\mathrm{H}\alpha$ profile in detail. 
The fitting results show that the redshift velocity typically has a value of around $50 \ \mathrm{km \ s^{-1}}$, similar to \citet{RN62}. We also confirm that the $\mathrm{H}\alpha$ EW of the Sun-as-a-star exhibits a similar trend to the GOES SXR in time, as reported by \citet{RN63}.

The Sun-as-a-star profiles we derived have a long-lasting red asymmetry persisting from the impulsive phase to the end of the flare. Similar results were also found in studies with high spatiotemporal resolution data \citep{2015ApJ...811..139T}. Three main factors are thought to be the cause of this phenomenon. First, the red asymmetry showing up during the impulsive phase is mostly connected to chromospheric condensation \citep{1985ApJ...289..414F}. Moreover, we observe that the rise in $\mathrm{H}\alpha$ EW occurs later than the rise in redshift velocity, which is also reported by \citet{RN62} and is believed to be a signature of chromospheric condensation \citep[e.g.,][]{1981SoPh...73..269L, RN42}. Second, the red asymmetry observed during the decay time of the flare may be associated with a ``warm rain" of the cooling of heated plasma as reported in \citet{2015ApJ...811..139T}. Third, the long duration of the red asymmetry could be caused by a superposition of unresolved events within the TR. In previous studies of solar flare utilizing the $\mathrm{H} \alpha$ waveband, a typical time for the persistence of red asymmetry is found to be around a couple of minutes \citep{RN42, 1995SoPh..158...81D}, both featured a spatial resolution of $2''$. Their long durations of red asymmetry are later associated with possible unresolved events within their spatial resolution \citep{2015ApJ...807L..22G}. In Sun-as-a-star studies, the effect of superposition can be more severe if phenomena such as a ``warm rain" take place inside a TR, thus leading to a red asymmetry with a longer duration than what is expected. 

In addition, we study the impact of the size of TR, which has long been a free parameter in the Sun-as-a-star method. 
Our results suggest that the size of TR influences the SNR of the Sun-as-a-star profiles, and a larger TR leads to a weaker flare signal. 
To analyze the impact of the size of TR on Sun-as-a-star profiles of the flare, we propose a new approach using DFT with Super Gaussian Window to denoise the Sun-as-a-star $\mathrm{H}\alpha$ contrast profiles. 
This approach helps remove the influence of irrelevant solar activities and observation errors, enabling us to focus on the impact of the size of TR. 
Our study finds that a larger TR leads to a decrease in the SNR of the Sun-as-a-star profile, which means that extracting spectral characteristics provided by the Sun-as-a-star profile becomes harder. 
We also find that there is an upper limit to the size of TR. 
If the signal of red asymmetry is too weak, it may not be possible to recognize it from the background noise. 
Therefore, in Sun-as-a-star studies, a smaller TR is preferred to extract the real signals.
From the stellar flare perspective, we can regard the TR as the distant star, and the flare inside the TR as the stellar flare. As we change the size of the TR, we equivalently simulate a series of stellar flares with different ratios of their areas to the stellar full disk. Our result indicates that, in stellar flare observations, for flares with relatively low intensity or small spacial scale, the signals regarding stellar dynamics may be undetectable. 
As the distance of a star gets further, the starlight may be more strongly absorbed by the gas and dust in the galaxy \citep{RN69, RN68}, and the observed intensity of flares from those stars will also decrease. 
We may infer from our result that, the red asymmetry in the line profiles of those stellar flare may become hard to be identified for the reason that they are too weak in intensity to be distiguished from the background noise. 

In the study, we analyze how the redshift velocity changes with the size of TR and how well the Sun-as-a-star method can capture this change. 
We use the three-component fitting method to calculate the redshift velocity for different Sun-as-a-star profiles with TRs of varying sizes. 
The results show that the redshift velocities calculated from different TRs during the same CHASE/HIS scan remain consistent. 
This suggests that the properties of chromosphere dynamics are well preserved despite the change in the size of the TR. 
Although it becomes more difficult to obtain useful spectral characteristics (i.e. redshift velocity, EW) about solar flares as the TR gets larger, the changes in the size of TRs will not significantly affect the detection of the redshift velocity. 
Therefore, from the perspective of redshift velocity, the Sun-as-a-star method allows for some flexibility in selecting TRs, as long as they cover the target flare. 

The time-averaged Sun-as-a-star profile is used to simulate the long-term exposure in stellar flare observations \citep{RN66, RN46}. 
By comparing the time-averaged profile with the time-resolved profile, we determine how exposure times affect the determination of the characteristics of stellar flare $\mathrm{H}\alpha$ spectra. 
After analyzing the profiles, we find that the $\mathrm{H}\alpha$ EW and Gaussian component EW no longer show a peak and both present a decrease in magnitude after being time-averaged. With the consideration that EW can measure the strength of the emission feature of the profile, the decrease in magnitude in EW can be interpreted as the decline in the overall intensity of the Sun-as-a-star $\mathrm{H}\alpha$ specrtum as well as the significance of the red asymmetry feature. 
This suggests that time-averaged profiles underestimate the flare characteristics in magnitude. 
Moreover, the emission intensities of stellar flares may also be underestimated under a long-term exposure. 

Note that we only investigate one solar flare in this study, and further investigations regarding the other solar activities' impact on $\mathrm{H}\alpha$ Sun-as-a-star profiles are under study based on CHASE/HIS data. 

\noindent \\ \hspace*{\fill} \\
We appreciate the referee for his/her comments and suggestions that improve our manuscript.
CHASE mission is supported by China National Space Administration. This study was supported by NSFC under grants 12333009 and 12127901, the National Key R\&D Program of China under
grant 2021YFA1600504, the CNSA project D050101, and Nanjing University’s National Innovation and Entrepreneurship project 202310284327Y.

\software{astropy \citep{astropy:2018}}

\bibliography{reference}{}
\bibliographystyle{aasjournal}

\end{document}